\documentclass[twocolumn,prb,aps,amsfonts,amssymb,amsmath,showpacs]{revtex4}
\usepackage{graphicx}
\usepackage{hyperref}

\begin{document}

\title{Casimir force measurements in Au-Au and Au-Si cavities at low temperature}

\author{J. Laurent, H. Sellier, A. Mosset, S. Huant, and J. Chevrier }
\affiliation{Institut N\'eel, CNRS et Universit\'e Joseph Fourier, B.P. 166, F-38042 Grenoble,
Cedex 9, France}

\date{\today}

\begin{abstract}

We report on measurements of the Casimir force in a sphere-plane geometry using a cryogenic force
microscope to move the force probe \textit{in situ} over different materials. We show how the
electrostatic environment of the interacting surfaces plays an important role in weak force
measurements and can overcome the Casimir force at large distance. After minimizing these parasitic
forces, we measure the Casimir force between a gold-coated sphere and either a gold-coated or a
heavily doped silicon surface in the 100--400\,nm distance range. We compare the experimental data
with theoretical predictions and discuss the consequence of a systematic error in the scanner
calibration on the agreement between experiment and theory. The relative force over the two
surfaces compares favorably with theory at short distance, showing that this Casimir force
experiment is sensitive to the dielectric properties of the interacting surfaces.

\end{abstract}



\maketitle

\section{Introduction}

Nanoelectromechanical systems (NEMS) are used in a broadening range of applications, such as
actuators, sensors, resonators, or modern nanocharacterization tools. Size reduction not only
allows for shrinking the energy consumption and for shortening the response time, but it also
allows integrating a broader range of functionalities on a single chip.~\cite{ekinci-2005} However,
quantum physics comes into play at the nanoscale and can affect NEMS behavior.~\cite{schwab-2005}
Understanding these effects, and possibly controlling them, is a necessary prerequisite to optimize
NEMS design. In turn, new applications driven by quantum effects can emerge, in particular in the
field of ultra-high-sensitivity force or displacement detection.

The Casimir force, discovered in 1948, is the archetypical force in this
framework.~\cite{casimir-1948} Its purely quantum origin results from the zero-point fluctuations
in the electromagnetic field. Since its theoretical prediction, the Casimir force has attracted the
interest of a large community of scientists ranging from cosmologists~\cite{antoniadis-2011} to
NEMS designers~\cite{lin-2005} through solid-state physicists. Experimentally, the first
confirmation~\cite{sparnaay-1958} of the Casimir effect was reported as early as 1958, but the
first quantitative study~\cite{lamoreaux-1997} of the Casimir force using a torsion pendulum was
reported not before 1997. Soon after, an important activity has been triggered thanks to the use of
atomic force microscopes (AFM)~\cite{mohideen-1998} or microelectromechanical systems
(MEMS).~\cite{chan-2001,decca-2003} Most of the experiments have been carried out with a cavity in
the sphere-plane geometry and very few in the plane-plane geometry~\cite{bressi-2002,antonini-2006}
that requires highly parallel surfaces.

The limited number of groups working on Casimir force measurements confirms how difficult these
experiments are.~\cite{ball-2007} This is explained by the small magnitude of this force as
compared to the electrostatic force, and by the stronger distance dependence scaling like the
inverse of the fourth (third) power of the distance in plane-plane (sphere-plane) geometry. In
order to check the validity of theories describing fundamental forces, the precision of the
experiments has been continuously improved, using metallic or semiconductor materials like
Au-Au,~\cite{lamoreaux-1997,chan-2001,decca-2005,decca-2007,kim-2008,vanzwol-2008,jourdan-2009,sushkov-2011}
Al-Al,~\cite{mohideen-1998,antonini-2006} Cr-Cr,~\cite{bressi-2002} Au-Cu,~\cite{decca-2003}
Au-Ge,~\cite{decca-2005b} Au-Si,~\cite{chen-2005,chen-2006,chen-2006b} Au-Si
grating,~\cite{chan-2008} Ge-Ge,~\cite{kim-2009} and Au-indium tin oxide.~\cite{deman-2009} In
parallel, the influences of layer thickness,~\cite{lambrecht-2000,genet-2003} surface
roughness,~\cite{palasantzas-2005} grating structure,~\cite{lambrecht-2008} and material
conductivity~\cite{duraffourg-2006,lambrecht-2007,dalvit-2008} have been studied theoretically to
provide models for comparison or stimulate new experiments.

In this paper, we report on a detailed study of the Casimir force between a gold-coated sphere and
a doped silicon substrate at liquid helium temperature (4.2\,K) and compare it {\it in situ} with
the case of a gold-coated surface. The use of the sphere-plane geometry avoids the challenge of
controlling with high accuracy the parallelism of two flat surfaces separated by a submicron gap.
Our aim is to reveal the dependence of the force on the materials properties and to compare it
quantitatively with theory. Thermalization at low temperature provides an exceptional mechanical
stability of the interacting surfaces, which is highly beneficial for long-term force measurements.
In principle, it should also improve the force sensitivity because of a reduced Brownian motion of
the cantilever, but other effects, such as optomechanical couplings, degrade the expected increase
in sensitivity in our experiment. Cooling down the experiment~\cite{decca-2010} also suppresses the
thermal contribution of the Casimir force,~\cite{genet-2000} allowing a direct comparison with the
zero-temperature theoretical calculations.

After a complete description of our calibration procedure, we show that parasitic forces can
perturb significantly the Casimir force measurements and that the setup environment can be modified
to suppress this artifact. We then discuss the variations of the minimizing potential with distance
by considering first the patch potential effect,~\cite{speake-2003} and then a simple electrostatic
model~\cite{hadjadj-2002} that reproduces the data. Finally, we present relative measurements of
the Casimir force in the 100--400\,nm distance range obtained by changing {\it in situ} the sample
surface from gold to silicon. The relative force between the two materials is in qualitative
agreement with theory, but the absolute values of the force show a systematic error with respect to
the theoretical predictions that are tentatively attributed to an aging of the scanner calibration.

The paper is organized as follows. Section II describes the homemade low-temperature force
microscope. Section III explains the force measurements and data analysis. Section IV discusses the
origin and suppression of a long-range residual force due to the electrostatic environment. Section
V presents the results on the minimizing potential and Casimir force obtained for gold and silicon
surfaces. Conclusions are drawn in Section VI.

\section{Low-temperature force microscope}

\begin{figure}
\begin{center}
\includegraphics[width=8.2cm]{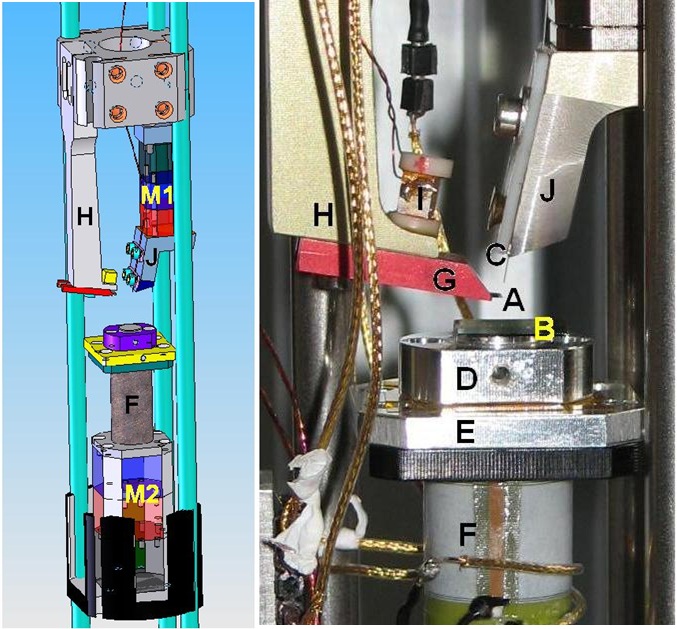}
\caption{Drawing and photograph of the microscope head at the bottom part of the cage structure.
The functional parts of the microscope, labeled by capital letters, are described in the text.}
\label{figure1}
\end{center}
\end{figure}

We developed a new force microscope working in a cryogenic environment at 4.2\,K, as an evolution
of the room-temperature instrument developed by Jourdan \textit{et al}.~\cite{jourdan-2009} The
structure of the low-temperature instrument takes the shape of a 50-mm-diameter and 120-cm-long
modular system based on a tubular cage.~\cite{brun-2001} This cage structure links the top of the
instrument (that bears all the electrical and optical connections) to the microscope head located
at the very bottom. The main parts of the microscope (marked by capital letters in
Fig.~\ref{figure1}) are described below.

The force probe is based on an AFM cantilever with a 40-$\mu$m-diameter polystyrene sphere fixed at
the extremity with standard epoxy glue (A). The sphere and cantilever are coated with gold (more
than 200\,nm on the sphere side and 80\,nm on the backside) to provide an electric contact to the
sphere, which is one of the Casimir mirror. The root-mean-square roughness of the gold surfaces is
around 3\,nm, as measured by AFM. The probes have typically a resonance frequency $f_0$ about
40\,kHz and a spring constant $k$ about 10\,N/m. The cantilever chip is glued with silver paint on
a holder (G) made of anodized aluminum and then fixed on a long holder (H). The cantilever is
mechanically excited at resonance by a piezoelectric dither (I). The sample (B) is mounted with
silver paint on a holder (D) that is separated from the piezoelectric $z$-scanner (F) by a grounded
aluminum plate (E) for electrostatic screening of the high voltages applied to the scanner. The
cantilever motion is measured with a compact optical detection compatible with the severe space
constraints of cryogenics,~\cite{rugar-1989} using the interferometric cavity formed by the
flexible cantilever and the extremity of a single-mode optical fiber (C) anchored to the holder
(J). The fiber is positioned above the end of the lever with a set of XYZ cryogenic inertial motors
(M1) and adjusted such as to obtain an interferometric cavity with good displacement sensitivity.
The sample is approached below the force probe with another set of motors (M2) and the scanner (F)
is used to finely tune the gap between the two surfaces. The scanner has been calibrated by
interferometry and the hysteresis has been determined for defined scanner extensions. It could be,
however, that this calibration slightly evolves in time after successive thermal cycles as
discussed later in the analysis of the results.

The microscope and the entire cage structure are sealed into a 2-in.-diameter stainless steel tube
evacuated to a secondary vacuum and flushed with helium gas. The tube is then filled with a low
pressure of helium exchange gas (10\,mbar at room temperature) and immersed in a liquid helium
cryostat. During cooling down, it is necessary to continuously readjust the optical cavity with the
M1 motors to compensate for thermal contractions.

Measurements at low temperature have the advantage to benefit from strongly reduced thermal drifts
and thermomechanical noises that usually limit the room-temperature experiments. For instance,
position drifts of about 1\,nm/min at 300\,K are found to be reduced to less than 1\,nm/h at 4\,K.
This is of particular importance in the present study because the Casimir force strongly depends on
distance. In the same way, the frequency drift of the cantilever resonance is strongly suppressed
from 3\,mHz/min at 300\,K down to a negligible value at 4\,K. Finally, another advantage of
cryogenic temperature is to strongly suppress the cantilever Brownian motion induced by
thermomechanical force fluctuations.

In such cryogenic conditions, the force detection sensitivity is essentially limited by the optical
readout of the cantilever position. The intensity fluctuations of the laser beam are here the main
source of noise, well above the noise of the photodiode and its amplifier. In particular,
optomechanical couplings like radiation pressure and photothermal stress convert this intensity
noise into cantilever displacement noise, called {\it backaction} noise. As a consequence, the
low-temperature force sensitivity is found to be of the same order as the room-temperature
sensitivity $\sqrt{S_{FF}}\approx 10\,\mathrm{fN}/\sqrt{\mathrm{Hz}}$. This situation could be
improved by a broadband stabilization of the laser beam intensity or by the coherent coupling of
laser noise and backaction noise, as demonstrated recently in our interferometric
setup.~\cite{laurent-2011} Therefore the only, but very rewarding, advantage of the low temperature
turns out to be the exceptional mechanical stability of the microscope.

These retarded optomechanical forces also modify the resonance frequency and damping rate of the
microlever through an optical spring effect induced by the interferometric
process.~\cite{braginsky-1970,braginsky-2002,favero-2009} Depending on the optical cavity detuning,
this effect can induce self-oscillations \cite{dorsel-1983,vogel-2003} or provide self-cooling of
the thermal noise.~\cite{metzger-2004,arcizet-2006,gigan-2006} The detection conditions have thus
been optimized by choosing the cooling side of the detuning and by adjusting the laser power.

\section{Data acquisition and calibration}

\begin{figure}
\begin{center}
\includegraphics[width=\columnwidth,clip,trim=0 0 0 0]{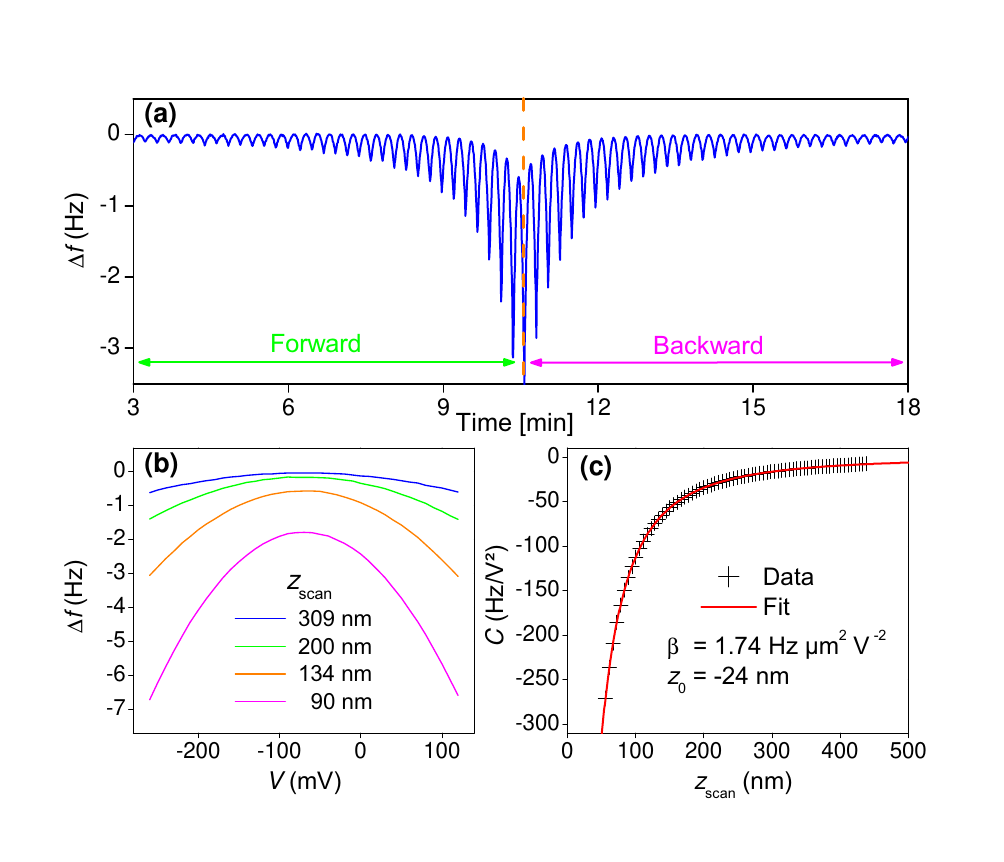}
\caption{(a)~Frequency shift $\Delta f$ of the cantilever resonance as a function of the time
elapsed during a forward (left) and backward (right) scan of the sample toward the sphere fixed on
the cantilever. Forward and backward scans are not perfectly symmetric because of the scanner
hysteresis. (b)~Examples of $\Delta f(V)$ parabolas recorded by sweeping the probe-surface bias $V$
for different scanner positions $z_\mathrm{scan}$. (c)~Fit of the parabola curvatures $C$ versus
scanner position $z_\mathrm{scan}$ used to obtain the force calibration factor $\beta$ and the
position $z_0$ of the contact.} \label{figure2}
\end{center}
\end{figure}

Instead of measuring directly the electrostatic or Casimir force $F(z)$ in static mode, we measure
its force gradient $G(z)=\frac{dF}{dz}$ in dynamic mode, which is given by the frequency shift
$\Delta f=-\frac{f_0}{2k}G$ of the cantilever resonance. The lever is excited with a piezoelectric
dither at its mechanical resonance and the lever vibration is measured by interferometry with the
optical fiber. The oscillation amplitude and phase are recorded with a lock-in and a phase-locked
loop tracks the resonance frequency $f$ when the probe is submitted to a force gradient. The
resonance frequency shift is defined by $\Delta f=f-f_0$, where $f_0$ is the free resonance
frequency.

Since the zero sphere-plane distance cannot be determined by bringing the sample into contact with
the sphere, which would irreversibly damage the gold coating of the surfaces, the absolute distance
is determined by electrostatic calibration. During a sequence of force measurements, the sample is
approached to the sphere by small steps, and for each scanner position $z_\mathrm{scan}$, the
resonance frequency shift $\Delta f$ is measured for different bias voltages $V$ applied to the
sample with respect to the grounded sphere [Fig.~\ref{figure2}(a)]. The voltage is varied typically
over $\pm 200$\,mV around the potential $V_\mathrm{min}$, which minimizes the electrostatic force
between the probe and sample. We obtain a series of $\Delta f(V)$ curves [Fig.~\ref{figure2}(b)],
which are fitted by the second-order polynomial:
\begin{equation}
 \Delta f = C (V-V_\mathrm{min})^2 + \Delta f_\mathrm{min}
 \label{equation1}
\end{equation}
where $C$, $V_\mathrm{min}$, and $\Delta f_\mathrm{min}$ are three adjustable parameters. The first
term on the right-hand side corresponds to the capacitive force, and the second term is the
frequency shift due to the remaining forces, including the Casimir force, obtained for
$V=V_\mathrm{min}$ at the summit of the parabola. The curvature coefficient $C$ is plotted as a
function of the scanner position $z_\mathrm{scan}$ and fitted with
\begin{equation}
 C=\frac{\beta}{(z_\mathrm{scan}-z_0)^2}
 \label{equation2}
\end{equation}
to obtain the sphere-sample contact position $z_0$ and the force probe calibration parameter
$\beta$ [Fig.~\ref{figure2}(c)]. The absolute distance between sphere and sample is then given by
$z=z_\mathrm{scan}-z_0$. We have additionally checked that the cantilever static deflection
generated by the electrostatic force and the Casimir force is negligible in the studied separation
range.~\cite{footnote}

The theoretical expression of the sphere-plane capacitive force gradient gives
$\beta=-\frac{f_0R\pi\epsilon_0}{2k}$, where $f_0$ is the free resonance frequency, $R$ the sphere
radius, $\epsilon_0$ the vacuum permittivity, and $k$ the cantilever stiffness. The experimental
value of $\beta$ extracted from the fit can therefore be used to transform the measured frequency
shift $\Delta f$ into a ``reduced force gradient'' $G/R$ without any other
parameter:~\cite{jourdan-2009}
\begin{equation}
 \frac{G}{R} = -\frac{2k}{f_0R} \Delta f = \frac{\pi\epsilon_0}{\beta} \Delta f
 \label{equation3}
\end{equation}
This electrostatic calibration of the force probe using the only parameter $\beta$ is more relevant
than the precise measurement of $R$ and $k$. The traditional determination of $k$ based on the
thermal noise spectral density and the equipartition theorem is indeed not possible at 4\,K because
of the dominant detection and backaction noises.~\cite{laurent-2011}

The measurement of $G/R$ in sphere-plane geometry allows a direct comparison between experiment and
theory within the so-called proximity force approximation (PFA):~\cite{derjaguin-1956}
\begin{equation}
 \frac{G}{R} \equiv \frac{1}{R} \frac{dF}{dz} = 2\pi\frac{F_{pp}}{A} \  \mathrm{for} \  z \ll R
 \label{equation4}
\end{equation}
where $F_{pp}/A$ is the force per unit area in plane-plane configuration, which is the quantity
usually calculated by theory.

The determination of the free resonance frequency $f_0$ is a difficult but important issue, since
it defines the origin of the frequency shift. This determination cannot be done when the sample is
further away from the probe than the scanner range (1.5\,$\mu$m), because using the step motor
would cause slight changes of $f_0$ due to mechanical vibrations that modify the system. In
practice, $f_0$ is determined just before starting the force measurements, at the maximum scanner
distance, and subsequently, we slightly adjust this value during the post-experimental analysis to
get a residual force going to zero at infinity. This small adjustment does not affect significantly
the data below 300\,nm.

\section{Suppression of the long-range residual force}

\begin{figure}
\begin{center}
\includegraphics[width=\columnwidth,clip,trim=0 0 0 0]{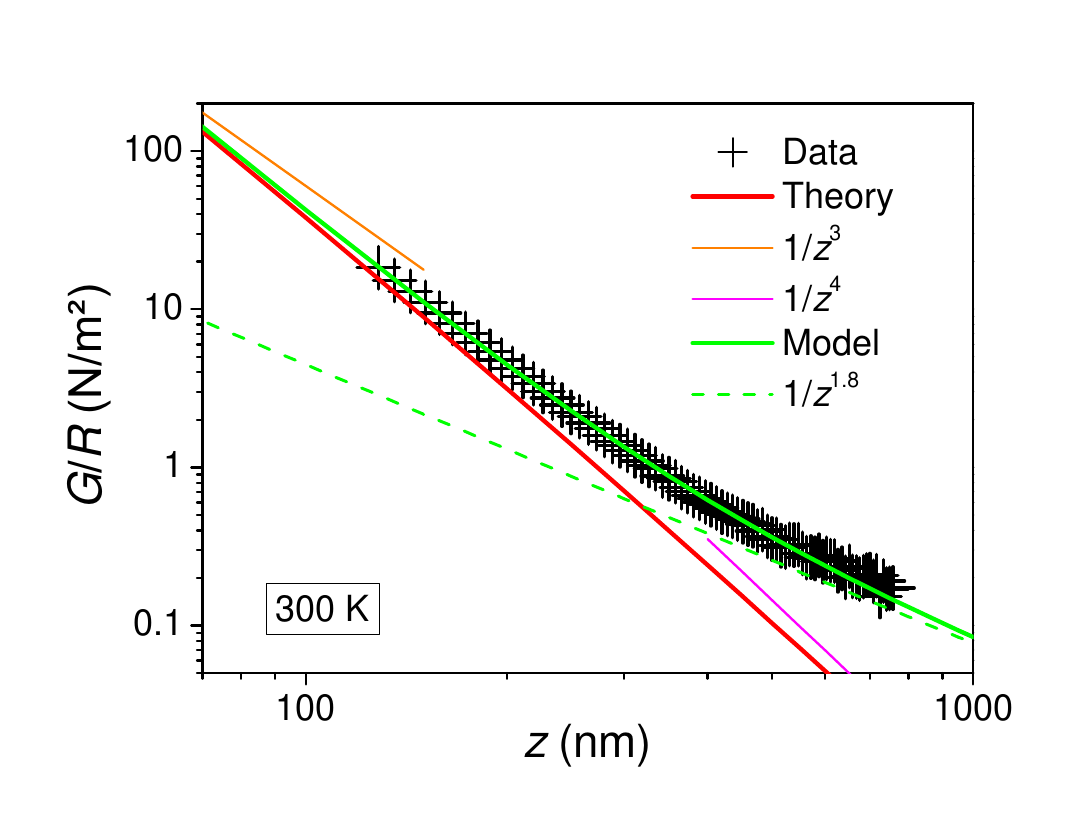}
\caption{Reduced force gradient at $V_\mathrm{min}$ versus distance between two gold surfaces in
sphere-plane geometry (measured at 300\,K). Experimental data are compared with the theoretical
prediction of the real Casimir force and with a model containing an additional long-range
contribution scaling like $1/z^{1.8}$.} \label{figure3}
\end{center}
\end{figure}

\begin{figure}
\begin{center}
\includegraphics[width=\columnwidth,clip,trim=0 0 0 0]{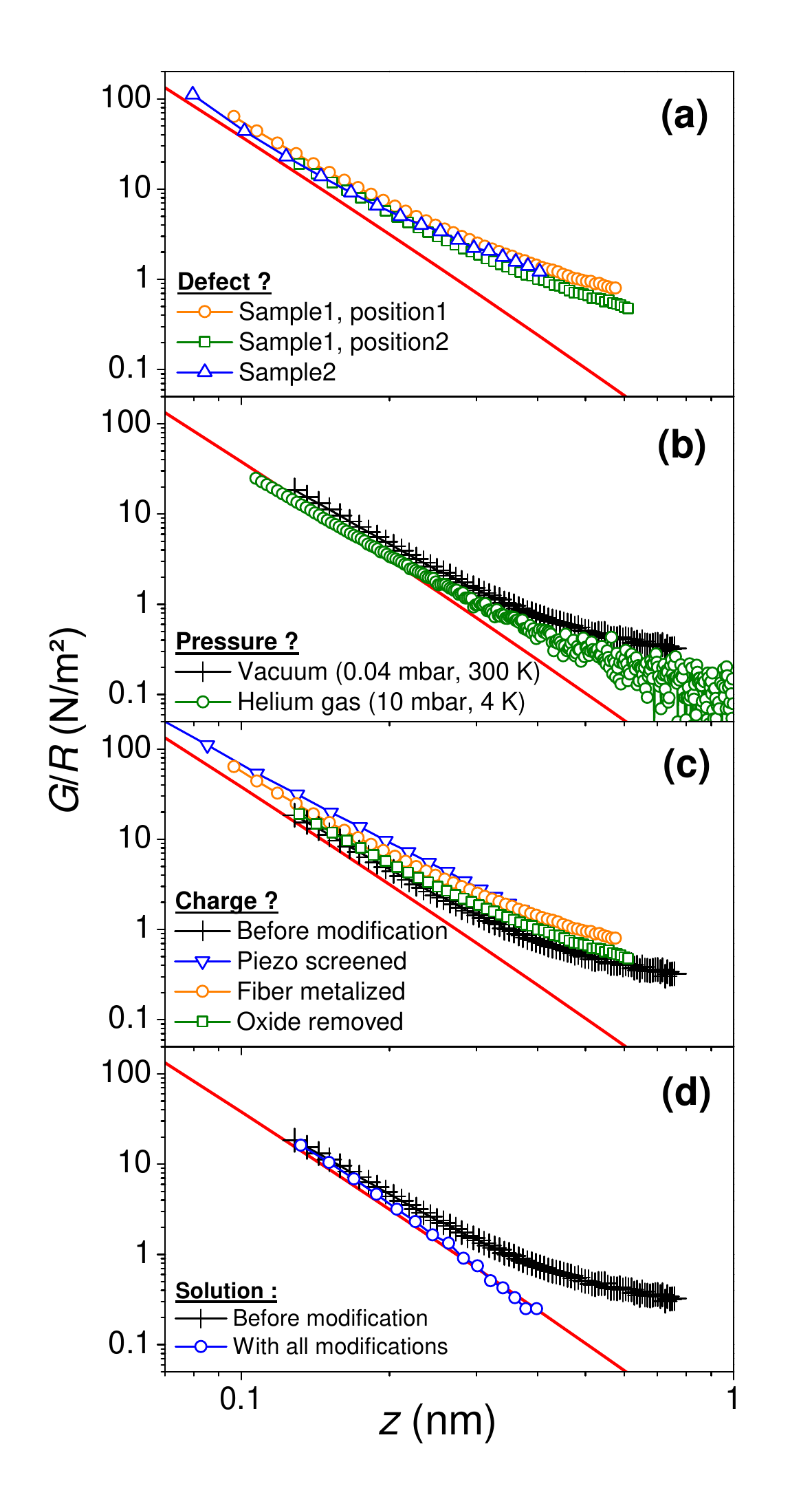}
\caption{Analysis of the long-range residual force at $V_\mathrm{min}$ showing the reduced force
gradient $G/R$ between two gold surfaces measured at 300\,K in vacuum: (a) at two positions located
1\,mm apart and on a different sample with the same probe; (b) at different temperatures and gas
pressures; (c) after a single change in the environment, either Faraday cages around piezoelectric
elements, or gold coating of the optical fiber, or removal of the oxide layer covering the anodized
chip holder; (d) with the three above modifications implemented together.} \label{figure4}
\end{center}
\end{figure}

At the minimizing potential $V_\mathrm{min}$, the residual frequency shift $\Delta f_\mathrm{min}$
should correspond \textit{a priori} to the searched-for Casimir force. The reduced force gradient
$G/R$ corresponding to $\Delta f_\mathrm{min}$ (measured at 300\,K) is plotted in
Fig.~\ref{figure3} as a function of the sphere-sample distance $z$ for two gold surfaces. The data
are compared with the theoretical prediction for Au-Au surfaces using the Drude
model.~\cite{lambrecht-2000} The force gradient in sphere-plane geometry is predicted to change
from the $1/z^3$ short-range regime (van der Waals) to the $1/z^4$ long-range regime (Casimir)
around the plasma wavelength of gold (136\,nm). It is clearly seen in the figure that the power law
of the experimental force gradient changes in a very different way, since the exponent is
decreasing with distance instead of increasing. There is obviously an additional ``parasitic''
force, which overcomes the Casimir force at large distance. By fitting the difference between data
and theory at large distance using a power law with the exponent as a free parameter, we obtain a
dependence scaling like $1/z^{1.8}$.

The origin of this parasitic force could be the inhomogeneity of the surface potential, which has
been first identified by Speake and Tenkel~\cite{speake-2003} as a source of residual electrostatic
force, which is not compensated at the minimizing potential $V_\mathrm{min}$. This inhomogeneity
originates from the random grain orientation of polycrystalline films, with different contact
potential on different crystal faces, or from an inhomogeneous layer of native oxide, adsorbed
contaminants, or chemical impurities. An experiment has reported such a long-range residual force
that could be attributed to this patch potential effect.~\cite{kim-2009,lamoreaux-2008,kim-2010}
Another experiment, however, with two aluminum surfaces, could not explain the observed additional
force by the spatial distribution of the contact potential that was measured directly by
Kelvin-probe force microscopy.~\cite{antonini-2009} In our case, the 1.8 power-law exponent is
larger than the value 1.44 obtained in Ref.~\onlinecite{kim-2009} (0.72 for force gives 1.44 for
force gradient), but is close to 2 as expected for a patch potential with small grains. In this
regime, the root-mean-square fluctuations of the gold contact potential in a granular film
($V_\mathrm{rms}=90$\,mV)~\cite{speake-2003} could be responsible for an electrostatic force
gradient as large as $G/R=\pi\epsilon_0 V_\mathrm{rms}^2/z^2=0.2$\,N/m$^2$ for $z=1\,\mu$m. This
order of magnitude is compatible with the long-range force visible in Fig.~\ref{figure3} above
300\,nm.

However, we discovered that this parasitic force could be suppressed after several modifications of
the measurement setup and is more probably due to the electrostatic environment of the force probe.
This conclusion is the result of a detailed analysis of many experiments carried out in different
situations as reported in Fig.~\ref{figure4}. First, we checked the reproducibility of this
parasitic force by comparing the results obtained at two different locations on the same gold
sample, and on a second gold sample [Fig.~\ref{figure4}(a)]. Only small differences are visible
between all three curves, with the same long-range residual force, therefore ruling out sample
specific artifacts, like defects or inhomogeneities. Then, we tested the influence of temperature
and exchange gas used for cooling the microscope head, because the gas confinement between the
sphere and surface could have produced an additional distance-dependent
dissipation.~\cite{siria-2009,drezet-2010,laurent-2011b} By comparing curves at 300\,K in vacuum
with curves at 4.2\,K in helium gas [Fig.~\ref{figure4}(b)], both showing the same additional
long-range component, we can rule out any significant effect of temperature and surrounding gas.
Note that the larger noise on the low-temperature data is the result of the optomechanical noise
discussed above in the paper. Finally, we analyzed the influence of the electrostatic environment
by testing separately a few changes to the setup, like covering the piezoelectric elements (scanner
and dither) with grounded Faraday cages, coating the cladding of the optical fiber with gold, or
removing the oxide layer of the anodized aluminum parts [Fig.~\ref{figure4}(c)]. Each change has
only a small impact on the parasitic force and none of them is able to cancel the parasitic force
alone. After implementing all three changes simultaneously, the parasitic force has finally
disappeared [Fig.~\ref{figure4}(d)]. We therefore conclude that the origin of this force was
probably not a patch potential effect, but more likely a force applied by residual charges in
different parts of the microscope head. The Casimir force measurements described in the next
section have been performed in these conditions with a clean electrostatic environment.

\section{Results for Gold-Gold and Gold-Silicon cavities}

\begin{figure}
\begin{center}
\includegraphics[width=8.2cm]{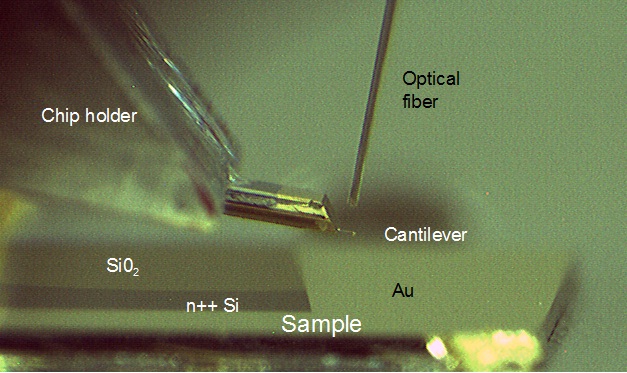}
\caption{Photograph showing the force probe, the optical fiber, and the sample made of a highly
doped silicon substrate partly covered by 150\,nm of gold (a third region is covered with a layer
of silicon dioxide).} \label{figure5}
\end{center}
\end{figure}

We now present our experimental results obtained at 4.2\,K with a gold-coated force probe on a
silicon substrate partly covered with 150\,nm of gold. The objective is to compare the Casimir
force gradient measured with the same sphere on two different materials. We compare a metal with a
semiconductor because these materials have very different electronic properties. The sample is made
of a heavily doped silicon substrate ($1.5\times 10^{19}$\,At/cm$^3$ phosphorus doping and
4.2\,m$\Omega$\,cm resistivity) in order to keep the surface conducting at low
temperature.~\cite{sze-2006} A region of the surface is then covered with 150\,nm of gold (e-beam
evaporation) with a sharp transition with the remaining part of the silicon substrate
(Fig.~\ref{figure5}). The translation stage (M2) is used to move the sample and place the selected
region in front of the sphere. The Casimir force can therefore be measured \textit{in situ} on the
two materials, using a single force probe in a single environment (gas and temperature), in order
to compare the data with better confidence than in separate experimental runs.

\subsection{Minimizing potential}

\begin{figure}
\begin{center}
\includegraphics[width=\columnwidth,clip,trim=0 0 0 0]{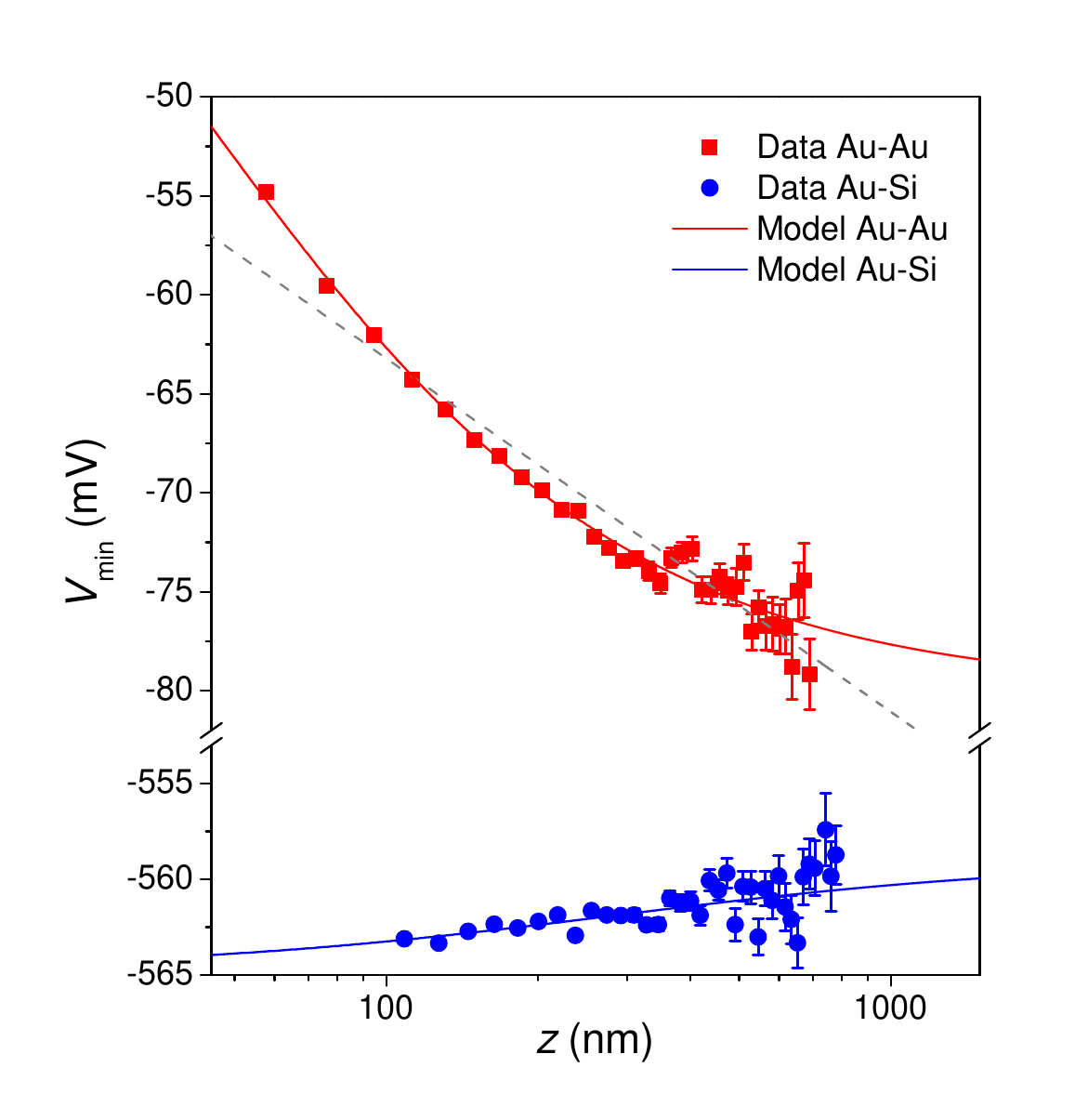}
\caption{Evolution with distance of the potential $V_\mathrm{min}$ applied at 4.2\,K on the sample
surface to minimize the electrostatic force on the Au-Au and Au-Si sphere-plane cavities.
Experimental data are fitted by the electrostatic model Eq.~(\ref{equation5}) (solid lines) and by
a logarithmic function (dashed line).} \label{figure6}
\end{center}
\end{figure}

The potential $V_\mathrm{min}$ applied on the sample to minimize the electrostatic force is plotted
in Fig.~\ref{figure6} as a function of the sphere-plane separation. These values are stable in time
and do not depend on the position along the surface. $V_\mathrm{min}$ is almost independent of the
distance for the silicon surface, around -560\,mV, but varies strongly with distance for the gold
surface, with an asymptote around -80\,mV at large distance. Since the contact potential $V_c$ is
expected to be zero for identical surfaces like in the Au-Au case, these results show that the
interpretation of $V_\mathrm{min}$ is more subtle.

Variations of the minimizing potential with distance have been observed previously in other Casimir
force experiments.~\cite{kim-2008,kim-2009,deman-2009} This effect can be explained by the
inhomogeneous surface potential (called patch potential) induced by the random distribution of
crystal orientations in gold films made of interconnected grains, several tens of nanometers in
diameter, with work function fluctuations $V_\mathrm{rms}\approx 90$\,mV.~\cite{speake-2003} When
the probe is close to the surface, the interaction area is small and more sensitive to the local
crystalline orientation, whereas at large distance, the interaction is averaged on a large number
of grains. Another explanation can be the presence of a smooth gradient of material work function
along the film.~\cite{kim-2010} In this context, the relation $V_\mathrm{min}(z)=a_1\log(z)+a_2$
was found to mimic the logarithm trend observed for two germanium
surfaces~\cite{kim-2009,lamoreaux-2008} and two gold surfaces.~\cite{deman-2009b} Here, the fit of
our Au-Au data with this relation (dashed line on Fig.~\ref{figure6}) is, however, not satisfactory
and we propose another model.

Casimir force experiments are not the only ones to evidence a distance dependence of the minimizing
potential. This effect is also observed in Kelvin-probe force microscopy and a model was developed
in this context by Hadjadj \textit{et al.},~\cite{hadjadj-2002} which takes into account the
interaction of the probe with its entire environment. By using a simple electrostatic model, these
authors found that the presence of metallic objects in the surrounding influences the minimizing
potential according to the relation
\begin{equation}
 V_\mathrm{min}(z) = V_c + \frac{b_1 z}{b_2 + z}
 \label{equation5}
\end{equation}
where $b_1$ and $b_2$ are related to the electrostatic potential and capacitance of the
environment, and $V_c$ is the contact potential obtained when $z$ tends to zero. By fitting our
Au-Au data with this model, as shown in Fig.~\ref{figure6}, we obtain $V_c=-20\pm 12$\,mV,
$b_1=-60\pm 12$\,mV, and $b_2=40\pm 13$\,nm. This simple electrostatic model reproduces very well
our experimental data and the contact potential is found very close to zero (considering the error
bar) as expected for two identical gold surfaces. This analysis demonstrates the influence played
by the environment of the force probe on the minimizing potential and shows that $V_\mathrm{min}$
can usually not be assimilated to the contact potential $V_c$ at finite distance.

The same analysis has been applied on the data obtained on the silicon surface and we obtain
$V_c=-565\pm 4$\,mV, $b_1=-6\pm 2$\,mV, and $b_2=300\pm 800$\,nm. Although we could have expected a
similar dependence on distance than for Au-Au because of the same environment, the minimizing
potential is found to be almost constant for Au-Si. An explanation might be that the sample has
been translated by a few millimeters to switch from gold to silicon, thereby slightly changing the
environment. The constant $V_\mathrm{min}$ for Au-Si confirms that the variations observed above
for Au-Au are not due to contact potential fluctuations, because we should also observe such
variations here, not due to the silicon surface, which is mono-crystalline, but due to the contact
potential fluctuations over the gold-coated sphere. The microstructure of the gold films could be,
however, different on the polystyrene sphere and on the silicon substrate, making a definite
conclusion difficult.

\subsection{Casimir force}

\begin{figure}
\begin{center}
\includegraphics[width=\columnwidth,clip,trim=0 0 0 0]{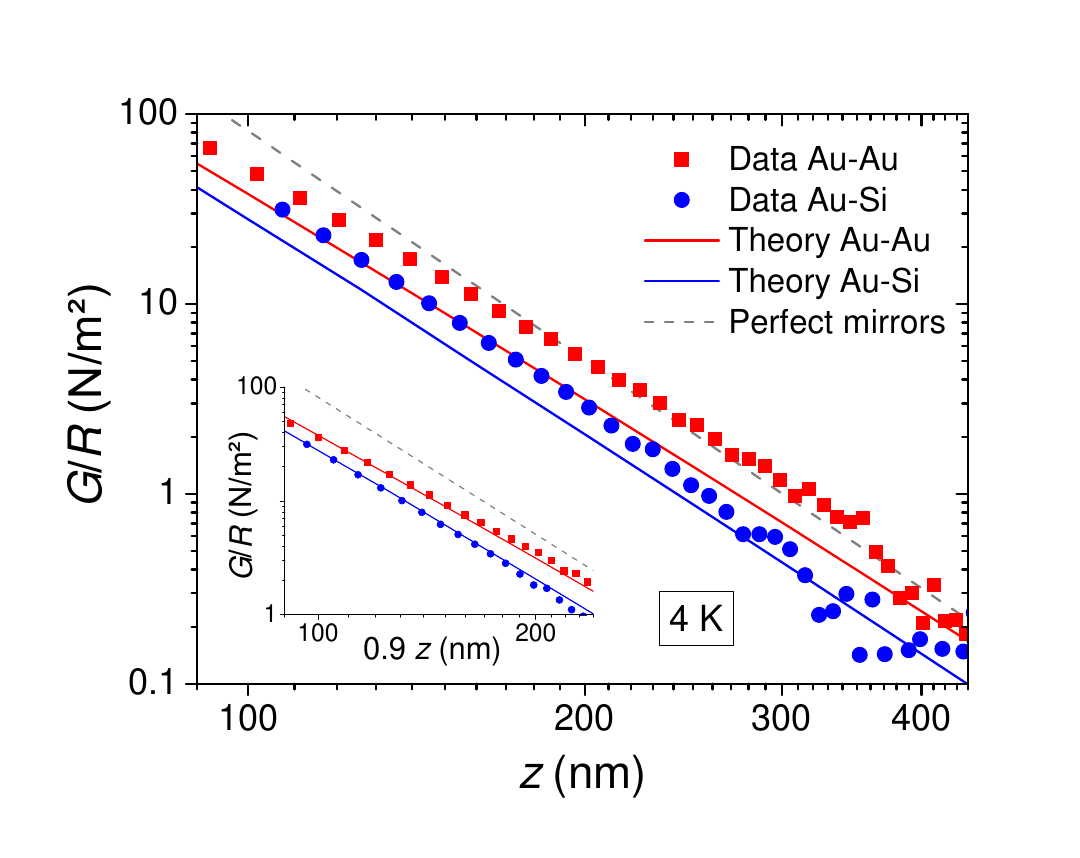}
\caption{Reduced Casimir force gradient $G/R$ versus distance $z$ measured at 4.2\,K between a gold
sphere and a gold surface or a doped silicon surface. Data are compared with theoretical
predictions for Au-Au and Au-Si cavities. The case of perfect mirrors is shown by a dashed line.
Inset: same data $G/R$ plotted for distances artificially reduced by a factor 0.9 to show that a
systematic error in the calibration might be the origin of the discrepancy between experiment and
theory.} \label{figure7}
\end{center}
\end{figure}

The Casimir force measured at 4.2\,K on the gold and silicon regions is presented in
Fig.~\ref{figure7} together with the theoretical predictions calculated for these specific
sample-probe configurations using the Drude model (see the
Appendix).~\cite{lambrecht-2000,duraffourg-2006,lambrecht-2007} It is clearly seen that the
measured Casimir force is weaker on doped silicon than on gold, as predicted by theory. The
experimental data are, however, above the theoretical curves by 50\%, i.e., much more than the
error in the force calibration factor $\beta$, which is better than 1\%. Recently, computations of
the Casimir force~\cite{pirozhenko-2006,svetovoy-2008} have emphasized the sensitivity of the
results to the choice of the materials optical data~\cite{ordal-1985,rakic-1998} used in the
calculations: for gold mirrors, the uncertainty is, however, only of about 5\%. The validity of the
PFA is another important assumption in the theory-experiment
comparison:~\cite{krause-2007,maianeto-2008} the error should be smaller than 1\% here since
$z/R<1\%$.~\cite{chen-2006} The large discrepancy between theory and experiment regarding the
absolute value of the force gradient requires, therefore, another explanation.

A possible source of error being the calibration of the scanner extension, we found that
multiplying the distance $z$ of each data point by a factor 0.9 translates the data points onto the
theoretical curves as shown in the inset of Fig.~\ref{figure7}. Since the force calibration is
dependent on the scanner calibration, it is in fact a factor 0.85 that should be applied on the
relative distance (before the determination of $\beta$) in order to shift the data onto the
theoretical curves. The piezoelectric $z$-scanner was calibrated by interferometry nine months
before the force measurements reported here and it could be that the scanner extension has been
progressively reduced after successive thermal cycles between 300 and 4.2\,K. Since this hypothesis
could not be checked at the time of the experiment, we stop here the discussion on the absolute
comparison between experiment and theory, and now discuss the relative value obtained between gold
and silicon surfaces.

\begin{figure}
\begin{center}
\includegraphics[width=\columnwidth,clip,trim=0 0 0 0]{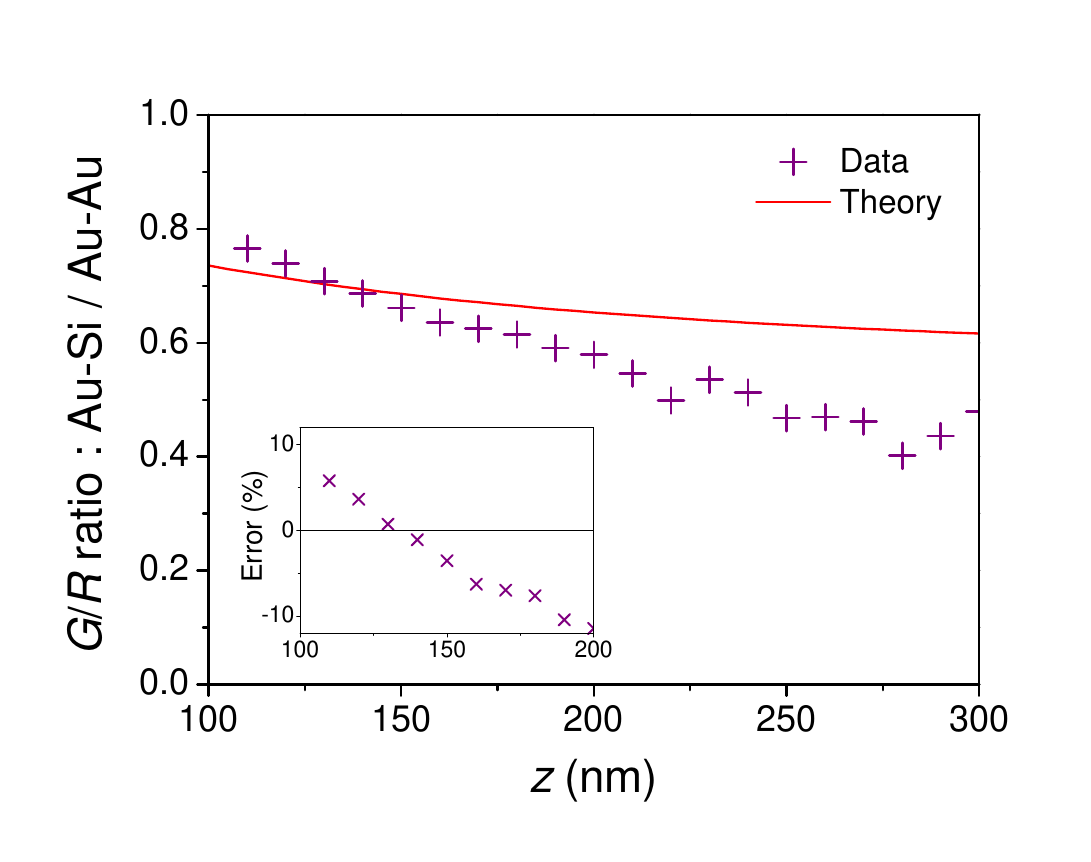}
\caption{Ratio between the force gradients measured for Au-Si and Au-Au cavities, compared with the
theoretical prediction. Inset: relative error between experiment and theory.} \label{figure8}
\end{center}
\end{figure}

The ratio of the Au-Si over the Au-Au force gradient is plotted in Fig.~\ref{figure8} for a series
of distances where the experimental force gradients have been determined by interpolation. The
ratio is lower than unity as expected for a cavity with a semiconductor plate, which is optically
less reflecting than gold. The ratio decreases progressively with distance as also expected from
theory,~\cite{lambrecht-2007} with a correction factor $\eta_F$, which saturates at large distance
to a lower value for Au-Si than for Au-Au. Quantitatively, the experimental ratio is of the same
order as the theoretical value at short distance, with an error less than 10\% in the 100--200\,nm
distance range, but the ratio decreases faster with distance than predicted by theory. These
results show that, although the absolute comparison with theory is not possible here, the material
dependence of the Casimir force is clearly evidenced when the surface is changed from gold to
silicon.

To improve this experiment in the future, the scanner extension should be measured by
interferometry {\it in situ} during the measurement to avoid the effect of thermal cycles on the
scanner piezoelectric coefficient. The detection sensitivity could be also improved by
stabilization of the laser intensity down to the shot noise level, in order to minimize the
detection and backaction noises, and take advantage of the strongly suppressed thermomechanical
noise at 4.2\,K.

\section{Conclusion}

From an instrumental point of view, we have shown that the presence of a long-range parasitic force
at the minimizing potential can be related to the electrostatic environment of the force probe.
Precise measurements of the Casimir force therefore require an accurate control of the environment,
like screening of every insulating part close to the probe: chip holder, optical fiber, and
piezoelectric actuators. Regarding the minimizing potential, the variations with distance observed
for the Au-Au cavity could be explained by a model taking into account the electrostatic potential
of the environment, and the absence of variation for the Au-Si cavity indicates that there is no
patch potential effect on the gold-coated sphere. Finally, we have shown that our \textit{in situ}
measurement of the Casimir force using a single spherical probe (gold) and two different surfaces
(gold or doped silicon) gives qualitatively the correct value for the relative force gradient,
although the absolute values are not correct due to a systematic error that might be attributed to
the scanner calibration.

The sensitivity of the Casimir force to material properties, as demonstrated here, could be used
for surface characterization in a new type of non-contact scanning force microscopy. Such a
technique would be the extension of the near-field van der Waals force microscopy to the retarded
Casimir regime at large separation. For a given force sensitivity, measuring at large distances
implies the use of micron-size spherical probes with a lower spatial resolution than the sharp tips
used in atomic force microscopy, but provides information on the optical properties of the
materials. With the Casimir force being obtained at the minimizing electrostatic potential, this
technique would be complementary to the measurement of the contact potential by Kelvin-probe force
microscopy.

\acknowledgments

The low-temperature AFM was designed and built by J-F. Motte. We thank T. Ouisse for an early
contribution to this work and for helpful discussion. This work has benefited from fruitful
discussions with P. Andreucci, L. Duraffourg, J-J. Greffet, G. Jourdan, and A. Lambrecht. The
samples and probes were prepared in the Nanofab clean-room facility. This research was supported by
the PNANO 2006 program of the Agence Nationale de la Recherche under the project name MONACO.

\appendix\section*{Appendix}

The Casimir force is computed for parallel plates of infinite thicknesses by taking into account
the real material properties as explained in Refs.~\onlinecite{duraffourg-2006,lambrecht-2007}. The
dielectric constant $\epsilon_r$ is modeled by a plasma frequency $\omega_p$ and a Drude relaxation
parameter $\gamma_p$, plus a Lorentz function with resonance frequency $\omega_0$, relaxation
parameter $\gamma_0$, and susceptibility $\chi_0$, describing interband transitions for gold and
intrinsic response for silicon:
\begin{equation}
 \epsilon_r(i\omega)=1+\frac{\omega_p^2}{\omega(\omega+\gamma_p)}+\frac{\chi_0\,\omega_0^2}{\omega^2+\omega_0^2+\omega\gamma_0}
 \label{equation6}
\end{equation}
The parameters used in the computation are listed in Table~\ref{table}. We checked that our
computation algorithm gives the correct results for the well-known Au-Au cavity, before computing
the force for our specific Au-Si cavity.
%
\begin{table}[b]
\begin{center}
\caption{Parameters of the dielectric function for gold (Au) and silicon (Si).} \label{table}
\begin{tabular}{|c|c|c|c|c|c|}
 \hline
  & $\omega_p$ & $\gamma_p$ & $\omega_0$ & $\gamma_0$ & $\chi_0$ \\
  & ($10^{15}$\,rad/s) & ($10^{15}$\,rad/s) & ($10^{15}$\,rad/s) & ($10^{15}$\,rad/s) & \\ \hline
 Au & 13.7 & 0.05 & 20 & 25 & 5 \\ \hline
 Si & 0.37 & 0.052 & 6.6 & 0 & 10.87 \\ \hline
\end{tabular}
\end{center}
\end{table}




\begin{thebibliography}{99}

\bibitem{ekinci-2005}
 K.L. Ekinci and M.L. Roukes, Rev. Sci. Instrum. \textbf{76}, 061101 (2005).

\bibitem{schwab-2005}
 K.C. Schwab and M.L Roukes, Phys. Today \textbf{58 (7)}, 36 (2005).

\bibitem{casimir-1948}
 H.B.G. Casimir, Proc. K. Ned. Akad. Wet. \textbf{51}, 793 (1948).

\bibitem{antoniadis-2011}
 I. Antoniadis, S. Baessler, M. Buchner, V.V. Fedorov, S. Hoedl, A. Lambrecht, V.V. Nesvizhevsky, G. Pignol, K.V. Protasov, S. Reynaud, Yu. Sobolev, C. R. Phys. \textbf{12}, 755 (2011).

\bibitem{lin-2005}
 W.H. Lin and Y.P. Zhao, Microsyst. Technol. \textbf{11}, 80 (2005).

\bibitem{sparnaay-1958}
 M.J. Sparnaay, Physica \textbf{24}, 751 (1958).

\bibitem{lamoreaux-1997}
 S.K. Lamoreaux, Phys. Rev. Lett. \textbf{78}, 5 (1997).

\bibitem{mohideen-1998}
 U. Mohideen and A. Roy, Phys. Rev. Lett. \textbf{81}, 4549 (1998).

\bibitem{chan-2001}
 H.B. Chan, V.A. Aksyuk, R.N. Kleiman, D.J. Bishop, and F. Capasso, Science \textbf{291}, 1941 (2001).

\bibitem{decca-2003}
 R.S. Decca, D. Lopez, E. Fischbach, and D.E. Krause, Phys. Rev. Lett. \textbf{91}, 050402 (2003).

\bibitem{bressi-2002}
 G. Bressi, G. Carugno, R. Onofrio, and G. Ruoso, Phys. Rev. Lett. \textbf{88}, 041804 (2002).

\bibitem{antonini-2006}
 P. Antonini, G. Bressi, G. Carugno, G. Galeazzi, G. Messineo, and G. Ruoso, New J. Phys. \textbf{8}, 239 (2006).

\bibitem{ball-2007}
 P. Ball, Nature \textbf{447}, 772 (2007).

\bibitem{decca-2005}
 R.S. Decca, D. Lopez, E. Fischbach, G.L. Klimchitskaya, D.E. Krause, and V.M. Mostepanenko, Ann. Phys. (N.Y.) \textbf{318}, 37 (2005).

\bibitem{decca-2007}
 R.S. Decca, D. Lopez, E. Fischbach, G.L. Klimchitskaya, D.E. Krause, and V.M. Mostepanenko, Phys. Rev. D \textbf{75}, 077101 (2007).

\bibitem{kim-2008}
 W.J. Kim, M. Brown-Hayes, D.A.R. Dalvit, J.H. Brownell, and R. Onofrio, Phys. Rev. A \textbf{78}, 020101 (R) (2008).

\bibitem{vanzwol-2008}
 P.J. van Zwol, G. Palasantzas, and J.Th.M. De Hosson, Phys. Rev. B \textbf{77}, 075412 (2008).

\bibitem{jourdan-2009}
 G. Jourdan, A. Lambrecht, F. Comin, and J. Chevrier, Europhys. Lett. \textbf{85}, 31001 (2009).

\bibitem{sushkov-2011}
 A.O. Sushkov, W.J. Kim, D.A.R. Dalvit, and S.K. Lamoreaux, Nature Phys. \textbf{7}, 230 (2011).

\bibitem{decca-2005b}
 R.S. Decca, D. Lopez, H.B. Chan, E. Fischbach, D.E. Krause, and C.R. Jamell, Phys. Rev. Lett. \textbf{94}, 240401 (2005).

\bibitem{chen-2005}
 F. Chen, U. Mohideen, G.L. Klimchitskaya, and V.M. Mostepanenko, Phys. Rev. A \textbf{72}, 020101(R) (2005).

\bibitem{chen-2006}
 F. Chen, U. Mohideen, G.L. Klimchitskaya, and V.M. Mostepanenko, Phys. Rev. A \textbf{74}, 022103 (2006).

\bibitem{chen-2006b}
 F. Chen, G.L. Klimchitskaya, V.M. Mostepanenko, and U. Mohideen, Phys. Rev. Lett. \textbf{97}, 170402 (2006).

\bibitem{chan-2008}
 H.B. Chan, Y. Bao, J. Zou, R.A. Cirelli, F. Klemens, W.M. Mansfield, and C.S. Pai, Phys. Rev. Lett. \textbf{101}, 030401 (2008).

\bibitem{kim-2009}
 W.J. Kim, A.O. Sushkov, D.A.R. Dalvit, and S.K. Lamoreaux, Phys. Rev. Lett. \textbf{103}, 060401 (2009).

\bibitem{deman-2009}
 S. de Man, K. Heeck, R.J. Wijngaarden, and D. Iannuzzi, Phys. Rev. Lett. \textbf{103}, 040402 (2009).

\bibitem{lambrecht-2000}
 A. Lambrecht and S. Reynaud, Eur. Phys. J. D \textbf{8}, 309 (2000).

\bibitem{genet-2003}
 C. Genet, A. Lambrecht, and S. Reynaud, Phys. Rev. A \textbf{67}, 043811 (2003).

\bibitem{palasantzas-2005}
 G. Palasantzas and J.Th.M. De Hosson, Phys. Rev. B \textbf{72}, 115426 (2005).

\bibitem{lambrecht-2008}
 A. Lambrecht and V.N. Marachevsky, Phys. Rev. Lett. \textbf{101}, 160403 (2008).

\bibitem{duraffourg-2006}
 L. Duraffourg and Ph. Andreucci, Phys. Lett. A \textbf{359}, 406 (2006).

\bibitem{lambrecht-2007}
 A. Lambrecht, I. Pirozhenko, L. Duraffourg, and P. Andreucci, Europhys. Lett. \textbf{77}, 44006 (2007).

\bibitem{dalvit-2008}
 D.A.R. Dalvit and S.K. Lamoreaux, Phys. Rev. Lett. \textbf{101}, 163203 (2008).

\bibitem{decca-2010}
 R.S. Decca, D. Lopez, and E. Osquiguil, Int. J. Mod. Phys. A \textbf{25}, 2223 (2010).

\bibitem{genet-2000}
 C. Genet, A. Lambrecht, and S. Reynaud, Phys. Rev. A, \textbf{62}, 012110 (2000).

\bibitem{speake-2003}
 C.C. Speake and C. Trenkel, Phys. Rev. Lett. \textbf{90}, 160403 (2003).

\bibitem{hadjadj-2002}
 A. Hadjadj, B. Equer, A. Beorchia, and P. Roca i Cabarrocas, Philos. Mag. B \textbf{82}, 1257 (2002).

\bibitem{brun-2001}
 M.~Brun, S.~Huant, J.C.~Woehl, J.-F.~Motte, L.~Marsal, and H.~Mariette, J. Microscopy  \textbf{202}, 202 (2001).

\bibitem{rugar-1989}
 D.~Rugar, H.J.~Mamin, and P.~Guethner, Appl. Phys. Lett. \textbf{55}, 2588 (1989).

\bibitem{laurent-2011}
 J. Laurent, A. Mosset, O. Arcizet, J. Chevrier, S. Huant, and H. Sellier, Phys. Rev. Lett. \textbf{107}, 050801 (2011).

\bibitem{braginsky-1970}
 V.B.~Braginsky, A.B.~Manukin, and M.Yu.~Tikhonov, Sov. Phys. JETP \textbf{31}, 829 (1970).

\bibitem{braginsky-2002}
 V.B.~Braginsky and S.P.~Vyatchanin, Phys. Lett. A \textbf{293}, 228 (2002).

\bibitem{favero-2009}
 I.~Favero and K.~Karrai, Nat. Photon. \textbf{3}, 201 (2009).

\bibitem{dorsel-1983}
 A.~Dorsel, J.D.~McCullen, P.~Meystre, E.~Vignes, and H.~Walther, Phys. Rev. Lett. \textbf{51}, 1550 (1983).

\bibitem{vogel-2003}
 M.~Vogel, C.~Mooser, K.~Karrai, and R.J.~Warburton, Appl. Phys. Lett. \textbf{83}, 1337 (2003).

\bibitem{metzger-2004}
 C.~Metzger and K.~Karrai, Nature (London) \textbf{432}, 1002 (2004).

\bibitem{arcizet-2006}
 O.~Arcizet, P.-F.~Cohadon, T.~Briant, M.~Pinard, and A.~Heidmann, Nature (London) \textbf{444}, 71 (2006).

\bibitem{gigan-2006}
 S.~Gigan, H.R.~B\"{o}hm, M.~Paternostro, F.~Blaser, G.~Langer, J.B.~Hertzberg, K.C.~Schwab, D.~B\"{a}uerle, M.~Aspelmeyer, and A.~Zeilinger, Nature (London) \textbf{444}, 67 (2006).

\bibitem{footnote}
The electrostatic force at $z=100$\,nm produces a static deflection of 2\,\AA\ for a cantilever
stiffness $k=8$\,N/m and a potential difference $V=500$\,mV.

\bibitem{derjaguin-1956}
 B.V. Derjaguin, I.I. Abrikosova, and E.M. Lifshitz, Q. Rev. Chem. Soc. \textbf{10}, 295 (1956).

\bibitem{lamoreaux-2008}
 S.K. Lamoreaux, e-print arXiv:0808.0885v2.

\bibitem{kim-2010}
 W.J. Kim, A.O. Sushkov, D.A.R. Dalvit, and S.K. Lamoreaux, Phys. Rev. A \textbf{81}, 022505 (2010).

\bibitem{antonini-2009}
 P. Antonini, G. Bimonte, G. Bressi, G. Carugno, G. Galeazzi, G. Messineo, and G. Ruoso, J. Phys.: Conf. Ser. \textbf{161}, 012006 (2009).

\bibitem{siria-2009}
 A. Siria, A. Drezet, F. Marchi, F. Comin, S. Huant, and J. Chevrier, Phys. Rev. Lett. \textbf{102}, 254503 (2009).

\bibitem{drezet-2010}
 A. Drezet, A. Siria, S. Huant, and J. Chevrier, Phys. Rev. E \textbf{81}, 046315 (2010).

\bibitem{laurent-2011b}
 J. Laurent, A. Drezet, H. Sellier, J. Chevrier, and S. Huant, Phys. Rev. Lett. \textbf{107}, 164501 (2011).

\bibitem{sze-2006}
 S.M. Sze and K.N. Kwok, \textit{Physics of Semiconductor Devices}, 3rd ed. (Wiley, New York, 2006).

\bibitem{deman-2009b}
 S. de Man, K. Heeck, and D. Iannuzzi, Phys. Rev. A \textbf{79}, 024102 (2009).

\bibitem{pirozhenko-2006}
 I. Pirozhenko, A. Lambrecht, and V.B. Svetovoy, New J. Phys. \textbf{8}, 238 (2006).

\bibitem{svetovoy-2008}
 V.B. Svetovoy, P.J. van Zwol, G. Palasantzas, and J.T.M. De Hosson, Phys. Rev. B \textbf{77}, 035439 (2008).

\bibitem{ordal-1985}
 M.A. Ordal, R.J. Bel, R.W. Alexander, Jr, L.L. Long, and M.R. Querry, Appl. Opt. \textbf{24}, 4493 (1985).

\bibitem{rakic-1998}
 A.D. Rakic, A.B. Djurisic, J.M. Elazar, and M.L. Majewski, Appl. Opt. \textbf{37}, 5271 (1998).

\bibitem{krause-2007}
 D.E. Krause, R.S. Decca, D. L\'opez, and E. Fischbach, Phys. Rev. Lett. \textbf{98}, 050403 (2007).

\bibitem{maianeto-2008}
 P.A. Maia Neto, A. Lambrecht, and S. Reynaud, Phys. Rev. A  \textbf{78}, 012115 (2008).

\end{thebibliography}
\end{document}